# Thermal and Magnetic Properties of the Spin-Chain Material, $Pr_3RuO_7$

M. Freamat, X.N. Lin, V. Durairaj, S. Chikara, G. Cao, and J.W. Brill

Department of Physics and Astronomy, University of Kentucky, Lexington, KY 40506-0055

ABSTRACT

We present measurements of the magnetic moment and specific heat of single crystals of insulating, $Pr_3RuO_7$, with octahedral $RuO_6$ and pseudo-cubic $PrO_8$ chains, separated by layers of 7-coordinate praseodymium ions. The susceptibility indicates that antiferromagnetic order sets in at $T_N = 54$ K, with polarization approximately along the chain direction, but the susceptibility anisotropy above $T_N$ indicates that the exchange interaction is strongly anisotropic, with competing ferromagnetic interactions. This competition presumably gives rise to the observed metamagnetic behavior for applied fields along the chain direction. The specific heat anomaly at $T_N$ is unusually mean-field in shape and roughly consistent with ordering of the moments of the ruthenium ions and the "chain" praseodymium ions, whereas the remaining interlayer praseodymium ions appear to be magnetically inert. A Schottky-like anomaly at low temperature, however, suggests that the "ordered" chain praseodymium ions are still sensitive to a crystal field.

## 1. Introduction

Ruthenium-based oxides provide an excellent opportunity to study novel electronic and magnetic states. In this paper, we report our experimental investigation of the magnetic and calorimetric properties of high-quality single-crystals of $Pr_3RuO_7$. This compound belongs to a class of ruthenates with formula $Ln_3RuO_7$ (Ln for lanthanides) with fluorite-related phases, either crystallizing in the *Cmcm* space group of the orthorhombic system or in a disordered fluorite-phase, depending on the size of the Ln cations relative to pentavalent cation Ru[1-4].

It was shown[5-7] that the Pr cation is large enough for $Pr_3RuO_7$ to crystallize in the ordered phase. Its crystalline structure is layered along *a*-axis, each *bc*-plane layer containing chains of corner-sharing $RuO_6$ octahedra zigzagging along *c*-axis, flanked by edge-sharing eightfold coordinated $PrO_8$ distorted cubes involving one third of the Pr ions. The other two thirds of the Pr ions, with sevenfold oxygen coordination, are between the layers. The layer separation is about 5.5 Å, significantly larger than the intra-layer chain separation of about 3.7 Å. This characterizes an almost one-dimensional configuration and it is one of the reasons these materials attracted attention, since their anisotropy could translate into specific magnetic and electronic properties due to the dependency of electron magnetic coupling on the crystal directions. For example, spin density waves[8], spin-Peierls transitions[9] or the Haldane-gap opening in an integer-spin Heisenberg antiferromagnet[10] are phenomena associated with low dimensional spin systems. Another reason for studying the $Pr_3RuO_7$ ruthenate is the presence of multiple magnetic constituents: the common highly oxidized $Ru^{5+}$ 4d cations (spin S=3/2) could experience spin interactions with the magnetically active $Pr^{3+}$ ions, in addition to the intra-chain coupling and the degeneracy removing interaction with the crystalline electric field.

Previous measurements on this material were on powdered samples. The present measurements were on single crystal samples for which, in principle, the lower density of sinks for vacancies (such as grain boundaries or dislocations) and resulting less clustered distribution should decrease the impurity contribution and hence favor the intrinsic features compared to polycrystalline samples.

Single crystals of $Pr_3RuO_7$ were synthesized by the floating zone technique using a commercial image furnace (NEC-SC II). The starting materials were $Pr_6O_{11}$ and $RuO_2$ powders with atomic ratio of 2.7:1. The mixture of $RuO_2$ and $Pr_6O_{11}$ was ground in a mortar and preheated at 900°C in air for 15hrs. The heated powder was then reground and formed into a rod 6mm in diameter and 70mm in length using hydrostatic pressure, which was sintered in air at 900°C for 15hrs. After these initial heat treatments, crystals were grown in the image furnace, under $O_2$ pressure of around 0.25 MPa. The single crystals obtained from this growth were characterized by powder X-ray diffraction and energy dispersive X-ray spectroscopy. All results indicate that the crystals are single phase with lattice parameters (a=10.9802 Å, b=7.3839 Å, c=7.5304 Å) consistent with published values[5,6].

## 2. Magnetic Properties

The magnetic properties of the $Pr_3RuO_7$ single crystals were measured using a Quantum Design MPMS LX 7T SQUID magnetometer, with the field aligned approximately parallel or perpendicular to the crystalline c-axis. The direction of the field in the ab-plane was not determined. The results shown below were for an irregularly shaped crystal of mass 0.8 mg; similar results were obtained for other samples.

Shown in Fig.1a is the magnetic susceptibility $\chi$ (defined as M/B in the low field limit) as a function of temperature for fields B=0.01 T parallel and perpendicular to the **c**-axis. The susceptibility is strongly anisotropic, with $\chi_{//} > \chi_{\perp}$ at high temperatures, indicating that the easy axis for the spin polarization is along the $RuO_6$ and $PrO_8$ chains. Also as expected, the two components of the susceptibility bracket that measured for powder samples[6,7]. There is a sharp peak in $\chi_{//}$ and cusp in $\chi_{\perp}$ at 54 K that defines the Neel temperature, $T_N$, i.e. the onset of antiferromagnetic order, with the magnetization of the sublattices again (approximately) along the chain direction. Note that while antiferromagnetic ordering at $T_N = 54$ K is consistent with that seen in polycrystalline samples[5-7], the data in Fig.1a shows no sign of a second, weaker magnetic ordering (~35 K) observed in the polycrystalline samples[6,7]. While the origin of the discrepancy is not entirely clear, it cannot be ruled out that this anomaly is not intrinsic to the material but due to crystalline defects affecting the spin structure.

Unconventional magnetic interactions are indicated by the high temperature susceptibility. For temperatures between 150 K and 350 K, the susceptibility is an excellent fit to the Curie-Weiss law [$\chi=\chi_o + C/(T-\theta_C)$], as shown in the inset to Figure 1a, where $(\chi-\chi_0)^{-1}$ vs. T is plotted. The fitting parameters are the temperature independent term, $\chi_o = 1.74$ memu•mol$^{-1}$ for the c-axis and 2.29 memu•mol$^{-1}$ for the perpendicular direction, the Curie temperature $\theta_C = +48$ K for the c-axis and -21 K for the perpendicular direction, and the Curie constant C = 4.21 emu•K•mol$^{-1}$ for both directions, corresponding to an effective moment, $\mu_{eff} = 5.81$ $\mu_B$, where $\mu_B$ is the Bohr magneton. It is remarkable that $\theta_C$ is negative for the perpendicular direction and positive (and larger) for the c-axis, as shown in the inset of Fig.1a, (note that polycrystalline samples have a small, positive $\theta_C$.[5–7]) indicating that the exchange coupling is antiferromagnetic for the perpendicular direction and ferromagnetic and twice as strong for the c-axis. Such strong exchange anisotropy is presumably due to spin-orbit coupling of the praseodymium ions.

The polycrystalline samples had somewhat larger Curie constants (5.4-6.0 emu•K•mol$^{-1}$) and negligible values of the temperature independent term ($\chi_0$). This is presumably due to the analysis, e.g. temperature ranges of the fits. For example, we note, as shown in the inset to Figure 1, that ($\chi - \chi_0$)$^{-1}$ is a much more linear function of temperature for T > 150 K than for the polycrystalline sample of Ref. [5].

In contrast, the increase in susceptibility at low temperatures does not fit a Curie-Weiss law for either direction. This increase and the small magnitude of the low-temperature susceptibility are discussed further below.

The magnetic anisotropy is also illustrated by the isothermal magnetization, M. Shown in Fig. 2 is M(B) at T=2 K for the c-axis and the perpendicular direction (i.e. that of Figure 1 for which the cusp in $\chi$ is observed). The most remarkable feature is that while $M_\perp$ remains paramagnetic, there is an abrupt rise in $M_\parallel$, which marks a metamagnetic transition at $B_c$=3.5 T, a reversal of the local spin directions that changes the magnetic state from antiferromagnetic to ferromagnetic via a first order transition. (For other directions in the ab-plane, behavior intermediate between $M_\parallel$ and $M_\perp$ are observed.) A metamagnetic transition occurs in an antiferromagnet that has both strong anisotropy and competing ferromagnetic interactions[11], as we observe. It is interesting, therefore, that La$_3$RuO$_7$, with only a single magnetic species, exhibits a spin-flop (i.e. rotation of the spin-sublattice polarization) rather than a metamagnetic transition in a magnetic field[4]. As shown in Figure 2, the saturation moment of our Pr$_3$RuO$_7$ crystal is $M_s$ = 4.8 $\mu_B$/f.u, at T = 2K; with increasing temperature, $M_s$ decreases slowly and the transition broadens.[7]

### 3. Specific Heat

For heat-capacity measurements, we used the ac-calorimetry technique with chopped light as the heating source[12]. Since the incident power is unknown, only relative values of the heat capacity were obtained, so the specific heat ($c_P$) was normalized using higher-temperature values obtained with differential scanning calorimetry[13]. In the 15-250 K temperature range, the measurements were performed on a 21 mg sample mounted on a thermocouple thermometer, while in the 3-20 K range the sample was mounted on a cernox bolometer. (Note that in both cases the heat capacity of the thermometer was much less than that of the sample.) The chopping frequency $\omega$ was tested and adjusted in different temperature intervals to be between the internal and external relaxation times. Typically, $\omega/2\pi \approx 4$ Hz for thermocouple based measurements and $\omega/2\pi \approx 10$ Hz in the temperature range using the bolometer. We did not observe any significant frequency dependence or hysteretic dependence on temperature. The inset to Figure 3 shows the specific heat, normalized to the gas constant R = 8.31 J•mol$^{-1}$•K$^{-1}$, as a function of temperature, with the normalizing DSC data. Note that at room temperature the specific heat is ~ 90% of its Dulong-Petit value (33R), as expected.

We also attempted measurements on the 0.8 mg sample of Figures 1-2, but the internal and external time constants were too close for quantitative measurements. The results, however, were qualitatively similar to those of the larger sample.

Figure 3 shows the specific heat between 20 K and 80 K; a large anomaly is observed at $T_N \approx 54$ K. As for the susceptibility, we do not observe a second anomaly at T = 35 K; whereas a $\Delta c_P \sim 0.7R$ anomaly was observed at 35 K by Harada and Hinatsu[6], our noise level gives $\Delta c_P <$ 0.03R at this temperature, again suggesting that the 35 K anomaly is caused by defects affecting the spin order in the polycrystalline material. Otherwise, our anomaly $T_N$ is very similar in shape and size to that of Reference [6]. In particular, the anomaly is surprisingly "mean-field" in shape, not only in being much steeper (i.e. "vertical") on its high temperature side than on its low-temperature side, but in only extrapolating to its background value (the smooth curve in Figure 3) at T ~ $T_N$/2. Of course, mean-field specific heat anomalies are unusual for Neel transitions in localized systems, so our observed anomaly, $\Delta c_P$ ~ 2R, should be considered an upper limit to its mean-field value.

The low temperature specific heat is shown in Figure 4. In the inset, we plot $c_P$ vs. T, while $c_P$/T vs. $T^2$ is plotted in the main figure. For temperatures between 12 K and 20 K, the specific heat can be well represented by $c_P = \gamma T + \beta T^3$, with slope $\beta$ ~ 1.2 mJ mol$^{-1}$ K$^{-4}$, as also observed by Zhou *et al*[7] for their powder sample. If entirely due to phonons, this would give a (per atom) Debye temperature of $\Theta_D$ ~ 300 K, but the $\beta T^3$ term presumably also has a contribution from antiferromagnetic magnons, which would increase $\Theta_D$. From the data above 12 K, one would also estimate $\gamma$ ~ 160 mJ mol$^{-1}$ K$^{-2}$, However, we also observe a plateau in $c_P$ at low temperature (see the inset), so that $c_P$/T starts increasing below 10 K; Zhou *et al*[7] observe a somewhat smaller increase (and no $\gamma T$ term). This upturn in $c_P$/T, which can be fit as a Schottky anomaly, can give rise to our apparent $\gamma T$ term, as discussed below.

**4. Discussion**

As mentioned above, the crystal structure features a zigzag chain of corner-sharing RuO$_6$ octehedra and a row of edged-sharing PrO$_8$ pseudocubes that alternately run along the c-axis[5,6]. Hence, the magnetic interaction along the c-axis is expected to be stronger than for other directions. Furthermore, the chain magnetic interactions may be stronger for the Ru d-electrons than for the Pr f-electrons, since the corner-shared octahedra, with ~180°-Ru-O-Ru bond angles, are much more favorable for superexchange interactions than the edge-shared pseudocubes, with ~90°-Pr-O-Pr bond angles, so the ruthenium magnetic interactions are expected to drive the transition. For example, it was found that doping on the ruthenium sites much more strongly affects the phase transition than doping on the praseodymium sites[7]. Of course, interchain magnetic coupling is also needed for the phase transition, and the chain Pr ions are expected to be sensitive to, and probably order in, the alternating exchange field established by ruthenium ordering. The magnetic interactions of the interlayer praseodymium ions, both to each other and the chains, are expected to be much weaker, since the ionic spacing is large and there are not oxygen ions appropriate as intermediaries for superexchange.[6]

A model of the magnetic ordering should account for the following features: i) the values of the saturated magnetization, $M_s$ and high-temperature Curie constant; ii) the (relatively small) susceptibility and increase in $c_P/T$ at low temperatures; iii) the mean-field anomaly in $c_P$ at $T_N$; . While a quantitative explanation of all these features is beyond the scope of this paper, below we suggest a model which qualitatively accounts for them.

The Curie constants for free $Pr^{3+}$ (J=4) and (orbitally quenched) $Ru^{5+}$ (S=3/2) ions are 1.56 and 1.87 emu•K•mol$^{-1}$, respectively, so the Curie constant expected for $Pr_3RuO_7$ would be 6.6 emu•K•mol$^{-1}$ if all the Pr moments are unquenched, much greater than the observed value of C = 4.2 emu•K•mol$^{-1}$ On the other hand, if the 7-coordinate crystal field of the interlayer praseodymium ions renders these into non-magnetic singlets so that only the chain Pr moments are active, the expected value would be 3.4 emu•K•mol$^{-1}$, much closer to the observed value. Alternatively, one might suppose that the orbital moments are quenched for all the Pr ions, which therefore have Curie constants 1.0 emu•K•mol$^{-1}$ (appropriate for S=1, L=0), giving a total Curie constant of 4.9 emu•K•mol$^{-1}$, also close to the observed value. However, the saturated magnetization above the metamagnetic transition, $M_s$ =4.80 $\mu_B$/f.u. is close to the value obtained assuming full orbital quenching of not only the $Ru^{5+}$ ions (3 $\mu_B$) but also the chain $Pr^{3+}$ ions (2 $\mu_B$), with no interlayer praseodymium moment.   Hence both properties imply that the interlayer praseodymium ions are in singlet states; the change from free J=4 moments above $T_N$ to orbitally quenched S=1 moments at high fields and low temperatures for the chain Pr ions may signal a subtle d-f coupling, perhaps facilitated by the ferromagnetic order above the metamagnetic transition.   Also note in Figure 2 that there appears to be an additional anomaly in M at B=3.9 T (absent in polycrystalline samples), suggesting that the spin polarization occurs in steps, e.g. the Ru and Pr spins may become ferromagnetically aligned separately.

This picture is also supported by the small value of the low-temperature susceptibility. While there is an upturn in $\chi$ at low temperatures, the low-temperature susceptibility is over an order of magnitude smaller than would be expected if Pr moments remain disordered at the transition.

The complexity of the spin structure and unconventional nature of the antiferromagnetism are expected to be born out by measurements of the susceptibility anisotropy in the ab-plane. In particular, in preliminary measurements we have observed that rather than a cusp, a small peak is observed at $T_N$ for most in-plane directions; i.e. there seems to be a unique "hard-axis" rather than "hard-plane", as for conventional antiferromagnets. A possible reason is that the alternating ruthenium and chain praseodymium moments may polarize in different directions. The in-plane anisotropy is being further investigated.

The value of $\chi_0$, the temperature independent term in the Curie-Weiss fits to the high temperature susceptibility is similar to that of highly correlated metallic ruthenates with the Ruddlesden-Popper structure and the apparent value of the linear specific heat coefficient, $\gamma$ ~

160 mJ mol$^{-1}$ K$^{-2}$, is even larger than the coefficient observed in these materials.[14] These values seem to suggest that there is a sizable density of states in Pr$_3$RuO$_7$, and its insulating character results localization due to strong scattering rather than a band or correlation gap. (For example, an unusually large γ is also observed in insulating Gd$_2$RuO$_5$.[15]) However, these values of χ$_o$ and γ give Wilson ratios of only 0.94 for the ab-plane and only 0.71 for the c-axis. These small values in what would presumably be a strongly correlated conductor suggest that both χ$_o$ and γ do not originate from a "hidden" Fermi surface, e.g. χ$_o$ may be a van Vleck contribution from the praseodymium ions and γ an artifact of the upturn in c$_P$/T at lower temperatures.

In particular, we have fit the low temperature specific heat to the Schottky expression for excitation of a two-level system: c$_P$ = βT$^3$ + ν g R (δ/T)$^2$exp(δ/T) / [1 + g exp(δ/T)]$^2$, where δ is the excitation energy, g is the ratio of the ground state degeneracy to that of the excited state, and ν is the number of excitations per formula unit, <u>without inclusion of a γT term</u>. Two such fits, discussed below, are shown in Figure 4; for all fits, the excitation energy δ ≈ 14 K. While the fits are rough, they show that Schottky-like anomalies at low-temperatures can give rise to an apparent γ at higher temperature.

We associate the Schottky anomaly with crystal field excitation of the praesodymiums. While Pr crystal field splittings are typically large (~ 100 K), the distortion of the PrO$_8$ pseudocubes will open small gaps between states which are degenerate for the undistorted cube. For example, for a cubic crystal field with large separation of ligands, appropriate for the PrO$_8$ chains,[5,6] the crystal field ground state of a J=4 ion is a Γ$_5$ triplet, consisting of a Kramer's doublet degenerate with a singlet.[16] Then the small value of δ might be associated with the splitting of this degeneracy due to the distorted cubic environment.[5,6] The doublet, in turn, will be split by the alternating exchange field established by the antiferromagnetic ruthenium ordering. The net result, as shown in the lower inset to Figure 4, may be a small splitting ($\delta$) between the "spin-up" member of the doublet and the singlet state. Then, as the exchange field becomes established at $T_N$, only ~ half the chain praseodymium ions will go into the spin-up state, with the others in the singlet, if $\delta \ll T$. As the temperature falls below $\delta$, the spin-up state will become fully occupied. (Note that this may also account for the increase in $\chi$ at low temperatures.)

The dashed curve in Figure 4 shows a fit with $g = 1$, consistent with our model, yielding $\beta = 1.2$ mJ•mol$^{-1}$•K$^{-4}$, $\delta = 14.3$ K, and $\nu \sim 1/2$, i.e. implying the existence of a supercell, which has not been observed. On the other hand, setting $\nu = 1$ (appropriate for the chain praesodymiums), yields the fit shown by the solid curve, with $\beta = 1.1$ mJ•mol$^{-1}$•K$^{-4}$, $\delta = 13.9$ K, but $g \sim 2$, i.e. a degenerate ground state, inconsistent with our model. This cannot be improved by inclusion of a $\gamma T$ term in the fit, as this decreases the magnitude of $\nu/g$. (Also note that much better $\gamma=0$ fits to the specific heat than those shown can be obtained with larger and unphysical values of $\nu$ and g.)

Despite these problems with the fits, we assume that, by including coupling between the Pr moments, the basic idea of our model, that the exchange field and small crystal field split the $\Gamma_5$ triplet as shown in the Figure, is correct, and we use these to discuss the specific heat anomaly at $T_N$. In particular, we assume that, because of the nearby singlet with $\delta \ll T_N$, the orbital degeneracy of the chain praesodymiums changes from three to two at the transition..

As described in Section 3, the specific heat anomaly is very mean-field in shape, implying that the interactions are much longer range than usual for a non-itinerant Neel transition. This may be a consequence of the Ru-Pr interchain interaction effectively coupling more distant ruthenium moments. The anomaly for a magnetic transition in mean-field theory is given by[17]

$\Delta c_{MF} = (5 R/2) \Sigma [(\Gamma_j^2 - 1) / (\Gamma_j^2 + 1)]$, where the sum is over different ordering moments and $\Gamma_j$ is

the ratio of degeneracies above and below the transition; i.e. $\Gamma \approx 3/2$ for the chain praesodymiums. Harada *et al*[2,3] have argued that, in related $Ln_3RuO_7$ lanthanide compounds, the crystal field also reduces the (S=3/2) ruthenium ground state to a Kramer's doublet, so we assume $\Gamma = 2$ for the ruthenium ions. Then the expected total specific heat anomaly is $\Delta c_{MF}$ = 2.5R, close to the observed value ($\Delta c_P \sim 2R$). While this concurrence may be fortuitous, given our approximations (e.g. in dealing with the praseodymium crystal field and neglect of spin-coupling above $T_N$), it certainly is consistent with our assumption that the interlayer praesodymiums do not participate in the transition, which would make the anomaly much larger.

We expect the entropy change at the transition to be $\Delta S = R \, \Sigma \, \ln(\Gamma_j) = 1.1 \, R$. Of course, estimating the entropy from the specific heat, $\Delta S = \int dT \, \Delta c_P/T$, depends critically on one's estimate of the specific heat background. Harada and Hinatsu[6] found a large transition entropy ($\approx$ 3R) by assuming that the background (phonon) specific heat was the same as that of $La_3NbO_7$, which, as they discuss, is a questionable estimate for such a high $T_N$. Instead, using the cubic baseline shown in Figure 3, we obtain the *rough* estimate $\Delta S \sim 0.5 \, R$, suggesting that the correct baseline is either considerably below our estimate or that, despite the mean-field shape, there is a considerable fluctuation region above $T_N$, as assumed in Reference [6].

In summary, we present measurements of the magnetic and thermal properties of crystals of the spin-chain compound, $Pr_3RuO_7$. The material undergoes a Neel transition at $T_N = 54$ K which is unusual in its degree of anisotropy and apparent mean-field behavior. Unlike for powder samples, no second anomaly at 35 K is observed. The anisotropy and competing interactions result in a metamagnetic transition for fields along the chain direction. Both the specific heat and magnetization suggest that the interchain praesodymiums are not magnetically active so that only the chain praseodymium ions order with the rutheniums at $T_N$. However, this

12the ratio of degeneracies above and below the transition; i.e. $\Gamma \approx 3/2$ for the chain praesodymiums. Harada *et al*[2,3] have argued that, in related $Ln_3RuO_7$ lanthanide compounds, the crystal field also reduces the (S=3/2) ruthenium ground state to a Kramer's doublet, so we assume $\Gamma = 2$ for the ruthenium ions. Then the expected total specific heat anomaly is $\Delta c_{MF}$ = 2.5R, close to the observed value ($\Delta c_P \sim 2R$). While this concurrence may be fortuitous, given our approximations (e.g. in dealing with the praseodymium crystal field and neglect of spin-coupling above $T_N$), it certainly is consistent with our assumption that the interlayer praesodymiums do not participate in the transition, which would make the anomaly much larger.

We expect the entropy change at the transition to be $\Delta S = R \, \Sigma \, \ln(\Gamma_j) = 1.1 \, R$. Of course, estimating the entropy from the specific heat, $\Delta S = \int dT \, \Delta c_P/T$, depends critically on one's estimate of the specific heat background. Harada and Hinatsu[6] found a large transition entropy ($\approx$ 3R) by assuming that the background (phonon) specific heat was the same as that of $La_3NbO_7$, which, as they discuss, is a questionable estimate for such a high $T_N$. Instead, using the cubic baseline shown in Figure 3, we obtain the *rough* estimate $\Delta S \sim 0.5 \, R$, suggesting that the correct baseline is either considerably below our estimate or that, despite the mean-field shape, there is a considerable fluctuation region above $T_N$, as assumed in Reference [6].

In summary, we present measurements of the magnetic and thermal properties of crystals of the spin-chain compound, $Pr_3RuO_7$. The material undergoes a Neel transition at $T_N = 54$ K which is unusual in its degree of anisotropy and apparent mean-field behavior. Unlike for powder samples, no second anomaly at 35 K is observed. The anisotropy and competing interactions result in a metamagnetic transition for fields along the chain direction. Both the specific heat and magnetization suggest that the interchain praesodymiums are not magnetically active so that only the chain praseodymium ions order with the rutheniums at $T_N$. However, this order is not complete in that the chain praesodymiums are still subject to a small ($\delta \sim 14$ K) crystal field excitation, resulting in a low-temperature Schottky anomaly in the specific heat.

We thank L.E. DeLong for very helpful discussions. This research was supported by the National Science Foundation. Grants # DMR-0100572, DMR-0240813, and DMR-0400938.

**Figure Captions**

**Fig.1.** The magnetic susceptibility $\chi$ (defined as M/B) as a function of temperature for $\chi_{\parallel}$ and $\chi_{\perp}$ at B=0.01 T. Inset: $\Delta\chi^{-1}$ vs. T where $\Delta\chi \equiv \chi-\chi_o$; note that dots are the data, and solid lines are guides to the eye.

**Fig.2.** The isothermal magnetization M(B) at T = 2 K for $M_{\parallel}$ and $M_{\perp}$.

**Fig.3.** The specific heat vs. temperature near $T_N$. The smooth curve shows the background estimated by fitting the specific heat away from the transition to a third order polynomial in T. Inset: The specific heat over the entire temperature range (solid curve) and the differential scanning calorimetry results (dashed curve).

**Fig.4.** The low-temperature specific heat, plotted as $c_P/T$ vs $T^2$, with the Schottky anomaly fits discussed in the text. The upper inset shows $c_P$ vs. T. The lower inset shows the proposed splitting of the $\Gamma_5$ triplet: the Kramer's doublet (arrows), with magnetic moments $\mu$, are Zeeman split by the exchange field ($B_{ex}$) while the non-magnetic singlet ("0") is split from these by the crystal field of the distorted cube, giving net splitting $\delta \approx 14$K.

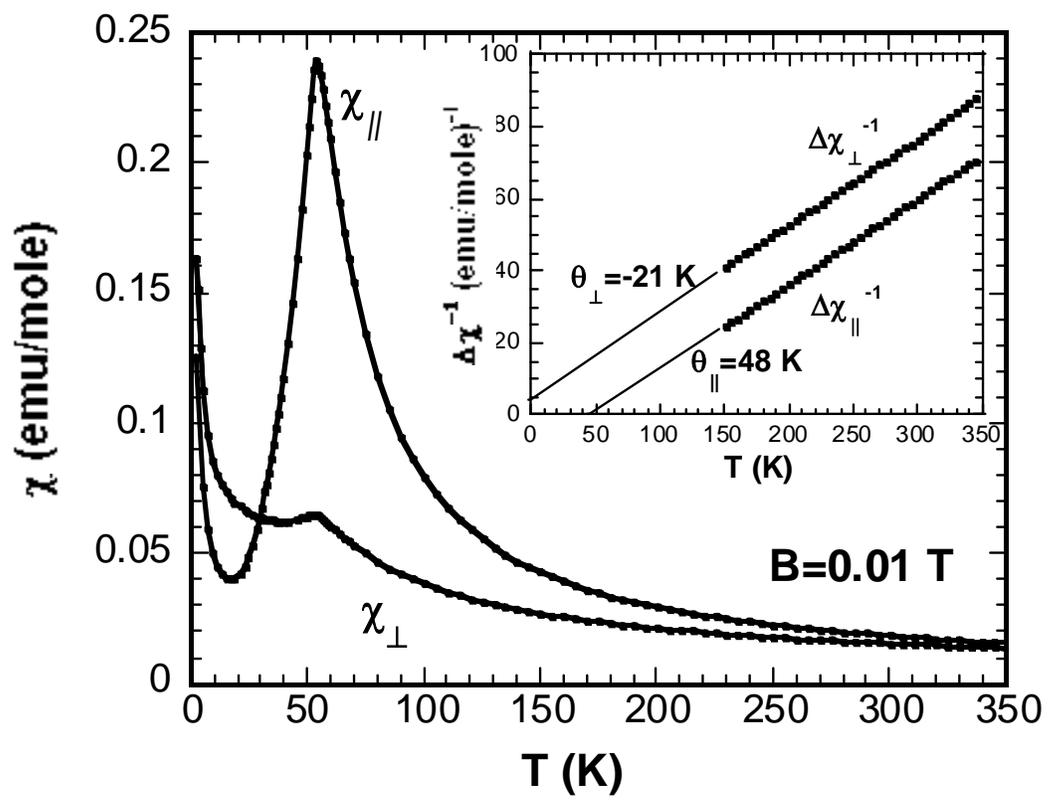

Figure 1

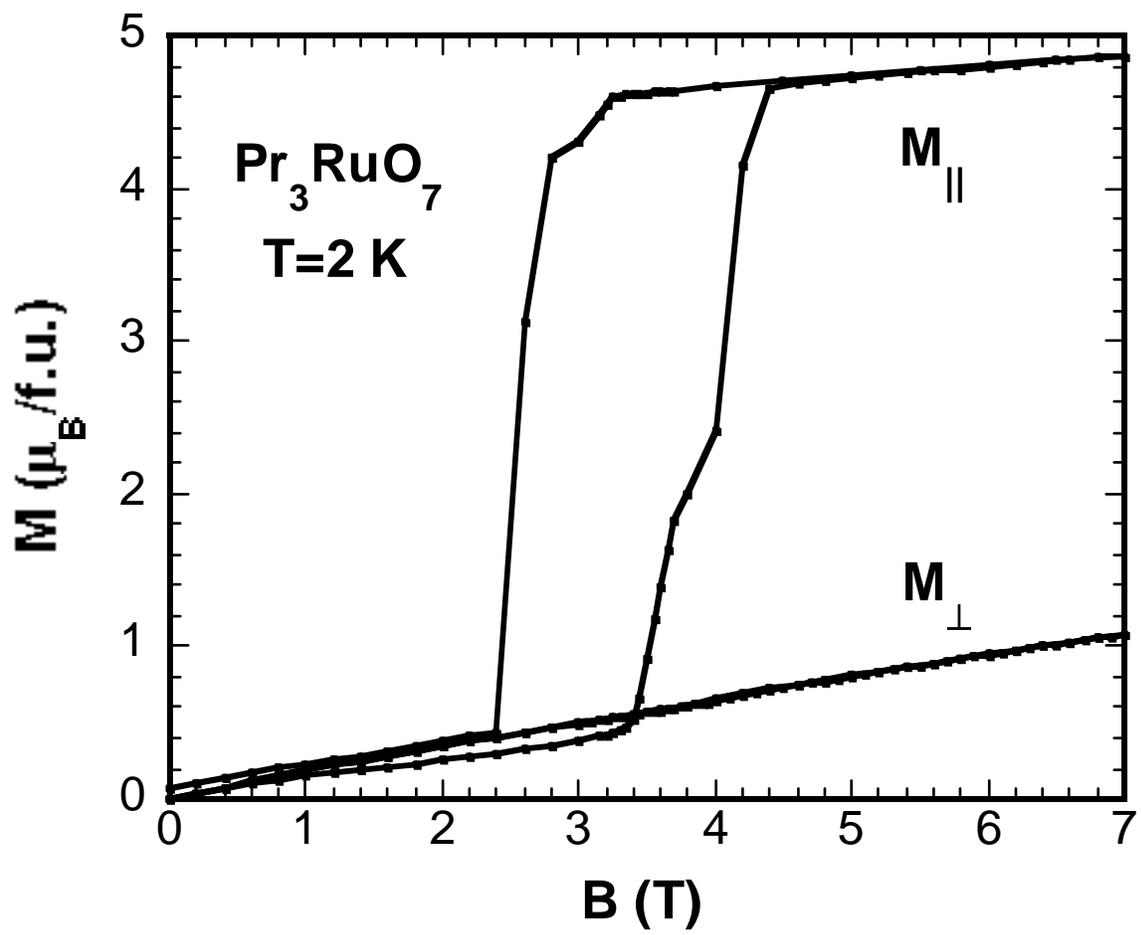

Figure 2

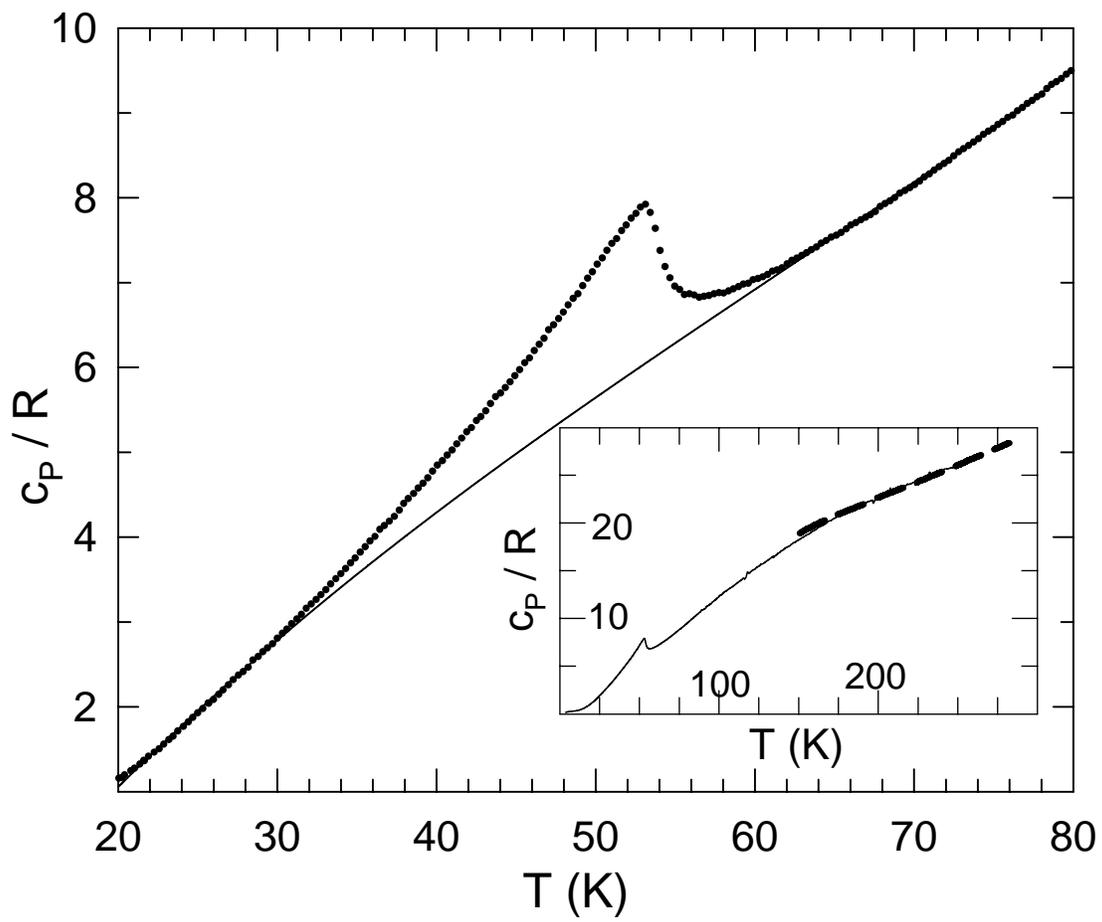

Figure 3

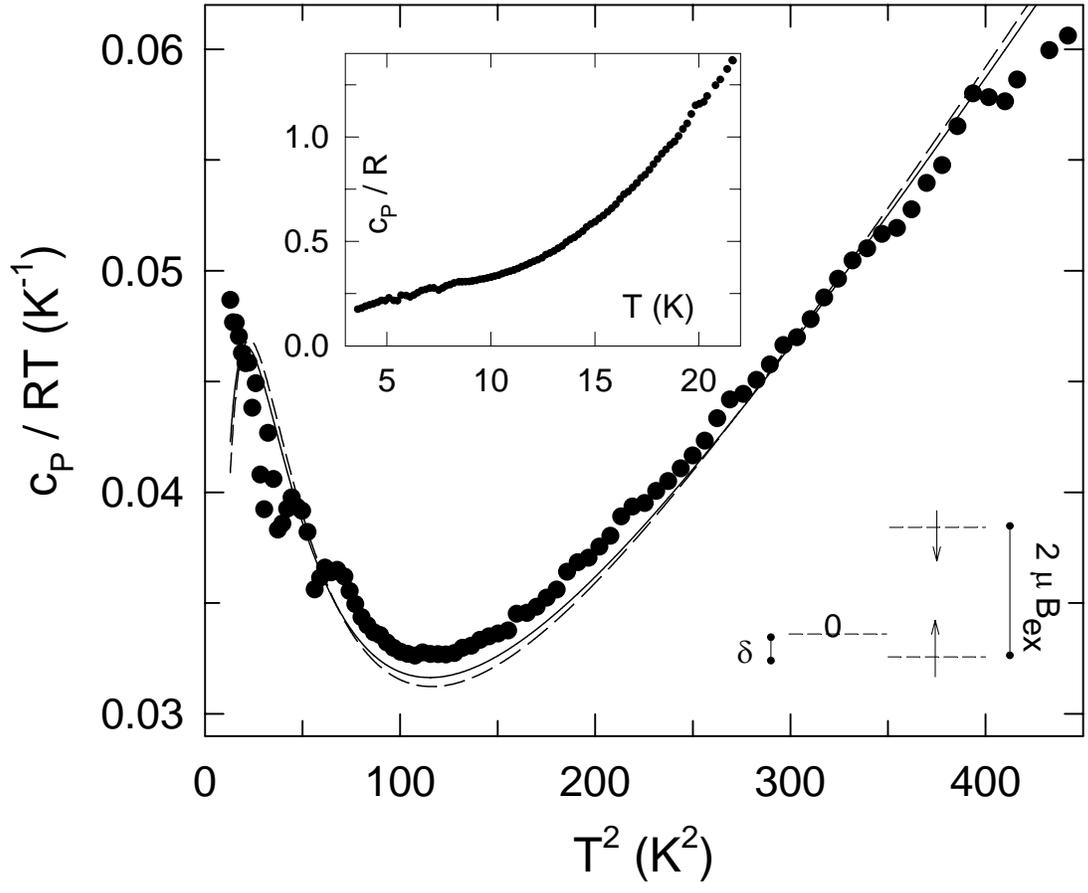

Figure 4